\documentclass[preprint,12pt,authoryear]{elsarticle}

\usepackage{amssymb}
\usepackage{amsmath}

\usepackage{lineno}
\usepackage{soul}
\usepackage{ragged2e}
\usepackage{float}
\usepackage{graphicx}
\usepackage{multirow}
\usepackage{subcaption}
\usepackage[
    colorlinks=true,    
    linkcolor=blue,     
    citecolor=red,      
    urlcolor=blue,      
    breaklinks=true     
]{hyperref}
\usepackage{hypcap}
\journal{Earth and Planetary Science Letters}

\begin{document}
\begin{frontmatter}



\title{Regime transitions of dynamos driven by fingering double-diffusive convection} 


\author[label1]{Wei Fan}
\author[label1,label2]{Yufeng Lin} 

\affiliation[label1]{organization={Department of Earth and Space Sciences, Southern University of Science and Technology},
            city={Shenzhen},
            postcode={518055}, 
            country={China}}
\affiliation[label2]{organization={Center for Complex Flows and Soft Matter Research, Southern University of Science and Technology},
            city={Shenzhen},
            postcode={518055}, 
            country={China}}

\begin{abstract}
Long-term cooling of planetary interiors is expected to progressively establish  thermally stable stratification in the liquid outer core, yet its influence on dynamo action remains incompletely understood. Here we present three-dimensional numerical simulations of fingering double-diffusive convection-driven dynamos to investigate how strengthening thermally stable stratification, modifies compositional convection driven flows and dynamos.
Our results show that dynamo evolution is governed by a competition between Lorentz-force regulation and stratification-induced flow reorganisation. When stratification is weak, the system remains in a strong-field regime characterised by a relatively stable, dipole-dominated magnetic field, despite reduced flow intensity. As stratification strengthens, the emergence and intensification of prograde equatorial zonal flow become the dominant manifestation of the systematic reorganization of the flow morphology.Once a critical stratification strength is exceeded, flow reorganisation becomes dominant, leading to an abrupt transition to a weak-field dynamo accompanied by a sharp decline in magnetic energy. In the weak-field regime, magnetic field intensity exhibits enhanced temporal variability and continues to decay as stratification further strengthens, eventually resulting in dynamo cessation even when compositional buoyancy persists. These results indicate that thermally stable stratification may play an important role in regulating planetary dynamos and provide new insight into the long-term magnetic evolution of terrestrial planets.
\end{abstract}

\begin{graphicalabstract}
\includegraphics[scale=0.6]{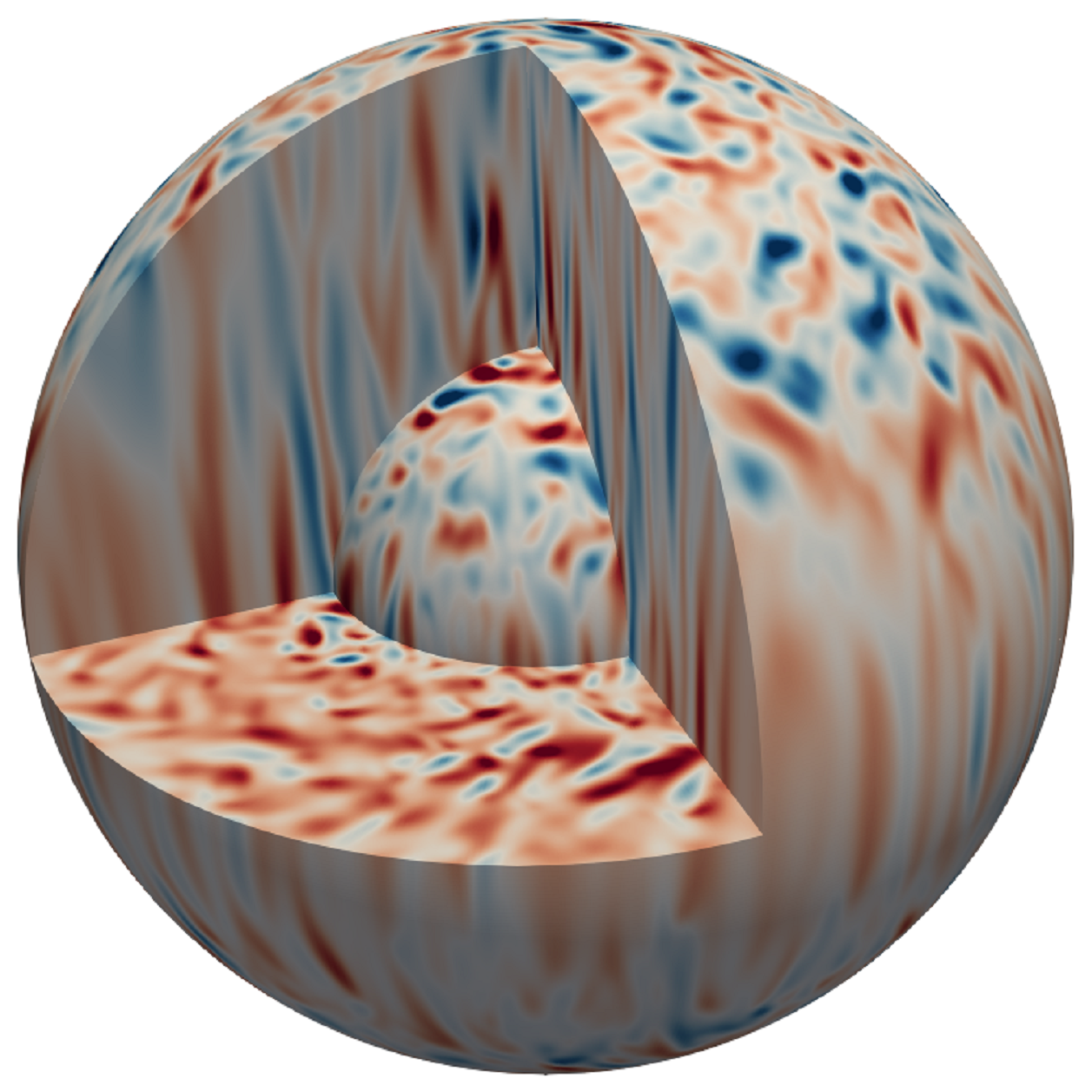}
\end{graphicalabstract}

\begin{highlights}
\item Thermally stable stratification induces an abrupt transition between strong and weak field dynamo regimes.
\item Enhanced thermally stable stratification systematically modifies flow morphology, leading to the emergence of prograde equatorial zonal flows.
\item Dynamo action can cease under strong thermally stable stratification, even in the presence of active compositional convection.
\end{highlights}


\end{frontmatter}




\section{Introduction}
\label{sec:intro}
The magnetic fields of terrestrial planets are widely believed to be generated by convective motions within their electrically conducting liquid cores, through a process known as the planetary dynamo \citep{Tikoo2022}. The nature and vigor of core convection are fundamentally controlled by a planet's long-term thermal and compositional evolution. In planetary cores, buoyancy driving the flow may originate from thermal and/or compositional sources \citep{Landeau2022sustaining,KNIBBE2018147,Andrewscience}. Thermal convection (TC) arises when heat extraction at the core-mantle boundary (CMB) and latent heat release during inner-core solidification maintain a superadiabatic temperature gradient, whereas compositional convection (CC) is driven by the release of light elements (e.g., silicon and oxygen) during inner-core crystallization \citep{hirose2013composition}. Notably, because the thermal diffusivity ($\kappa_T$) is much larger than the compositional diffusivity ($\kappa_C$), the interaction between these two buoyancy sources can give rise to a multiscale flow regime known as double-diffusive convection (DDC), with important implications for core dynamics and magnetic field generation.

As planetary cores cool over geological timescales, the heat flux transported through the core may decrease. If the heat flux at the CMB falls below the adiabatic value, a superadiabatic temperature gradient can no longer be maintained, leading to the formation of a thermally stable stratification (TSS) within the core \citep{DRISCOLL201436}, while compositional buoyancy may still sustain convection. Under such thermally stable but compositionally unstable conditions, the flow can organise into narrow, finger-like structures, forming a regime known as fingering double-diffusive convection (FDDC) \citep{Stern1960,Turner1985}. This mechanism is analogous to salt-finger convection in Earth's oceans, but its implications for planetary core dynamics and magnetic field generation may be far-reaching. FDDC has been proposed to play a key role in the liquid core of Mercury \citep{manglik2010dynamo,takahashi2019mercury}, where a thermally stable layer may help explain its weak and predominantly axisymmetric magnetic field \citep{Johnson2018}. Similarly, the cessation of the dynamo in Mars has been linked to the development of TSS in its core \citep{Greenwood2021,Hsiehsciadv}. These considerations suggest that FDDC may represent an important convective regime during specific stages of planetary evolution, with significant consequences for dynamo action.

FDDC has been extensively studied in oceanographic contexts \citep{radko2013double}, where it arises from the contrasting effects of temperature and salinity gradients on seawater density. Most existing studies have focused on non-rotating systems and have relied on local Cartesian models for theoretical analysis and numerical simulations \citep[e.g.][]{simeonov2008double,Ouillon_Edel_Garaud_Meiburg_2020,Li_Yang_2024}. While these studies have provided important insights into the underlying mechanisms of FDDC, their applicability to planetary cores remains limited. In planetary interiors, the combined effects of rapid rotation, spherical geometry, and magnetic field generation fundamentally shape flow dynamics. Numerical investigations of FDDC in rotating spherical shells are therefore essential for assessing its relevance to planetary dynamos. To date, however, most such studies have focused on the ``top-heavy'' regime, in which both thermal and compositional gradients are destabilising \citep{breuer2010thermochemically,trumper2012numerical}, as well as on the associated dynamo processes \citep{takahashi2014double,tassin2021geomagnetic,fan2025}. By contrast, the fingering regime, which is expected to arise naturally as TSS develops during planetary cooling, has received little attention in global rotating models.  Although stable stratification has been incorporated in some previous dynamo studies to mimic more realistic core conditions, it is often introduced indirectly through modifications of the background conductive temperature profile, rather than by solving the full set of double-diffusive governing equations \citep[e.g.][]{Wulff2025,STANLEY2008179,Gastinestable}. Such approaches cannot fully capture the fundamental role played by the disparity between thermal and compositional diffusivities. Therefore, the nonlinear evolution of FDDC and its influence on magnetic field generation in planetary core environments have not yet been systematically investigated in self-consistent, rotating dynamo models.

Despite the lack of systematic investigations, a few hydrodynamic studies of FDDC have shown that, under highly supercritical conditions, the presence of a moderate TSS can reduce the overall convective vigor while simultaneously enhancing large-scale zonal flows \citep{monville2019rotating,guervilly2022fingering,GRAY2026107570}. In the context of numerical dynamo simulations, \citet{takahashi2019mercury} showed that introducing TSS can reproduce several key features of the magnetic field of Mercury, including its weak intensity and spatial structure. However, this result was later shown to be highly sensitive to parameter choices \citep{Kolhey2025}, indicating that dynamo behavior in the FDDC regime depends on complex parameter interactions and nonlinear feedbacks. By contrast, \citet{mather2021regimes} reported no self-sustained dynamo solutions in FDDC models. Overall, the dynamical properties and magnetic field generation capability of FDDC in planetary-scale rotating systems remain poorly constrained, leaving significant scope for further investigation.

To address these uncertainties, we perform a series of three-dimensional numerical simulations of FDDC convection in a rotating spherical shell to investigate its ability to drive planetary dynamos. Our primary objective is to assess how dynamo behavior, including flow structure, magnetic field morphology, and field intensity, evolves as TSS strengthens, mimicking long-term planetary cooling. Compared with previous studies, our model employs a relatively large Lewis number, $Le = \kappa_T / \kappa_C = 100$, reflecting the strong disparity between thermal and compositional diffusivities expected in planetary cores. Although the exact diffusivity values remain uncertain \citep[e.g.][]{loper1981study,braginsky1995equations}, it is generally accepted that thermal diffusivity exceeds compositional diffusivity by several orders of magnitude, leading to estimated Lewis numbers in the range $10^2\,-\,10^4$ \citep{BOUFFARD2017552,wicht2019advances}. Numerical simulations at such large $Le$ are computationally challenging because of the wide separation of spatial and temporal scales \citep{BOUFFARD2017552}, and most previous studies have therefore adopted $Le=10$ to balance numerical feasibility with physical realism. By extending the parameter space to higher Lewis numbers, our simulations provide a closer approximation to planetary core conditions and enable a more robust assessment of the role of FDDC in planetary dynamo.

The remainder of this paper is organized as follows. Section \ref{sec:methods} introduces the numerical model and computational methods. Section \ref{sec:results} presents and discusses the simulation results, focusing on the effects of increasing TSS strength on flow and magnetic field behavior. Section \ref{sec:discussion} provides a general discussion. Finally, Section \ref{conclusions} summarizes the main conclusions.

\section{Methods}
\label{sec:methods}
\subsection{Governing Equations}
\label{sec:eq}
We consider Boussinesq convection of a homogeneous, electrically conducting fluid in a rotating spherical shell with inner radius $r_i$ and outer radius $r_o$, rotating uniformly at angular velocity $\boldsymbol \Omega=\Omega \boldsymbol{\hat z}$. FDDC is considered in a configuration where the imposed temperature difference $\Delta T = T_i - T_o$ establishes a TSS, while the imposed compositional difference $\Delta C = C_i - C_o$ is destabilising and provides the buoyancy driving for convection. The system is subject to a gravity field $\boldsymbol{g} = - g_0 \boldsymbol{r} / r_o$, where $g_0$ denotes the gravitational acceleration at the outer boundary. The magnetic field is matched to a potential field outside the fluid domain, with both the inner core and the mantle treated as electrically insulating. The shell thickness $D = r_{o} - r_{i}$ is used as the characteristic length scale,  $\Omega^{-1}$ as the time scale, $\Delta T$ and $\Delta C$ as the temperature and compositional scales, respectively, and $\sqrt{\rho \mu \eta \Omega}$ as the magnetic scale, where $\rho$ is the fluid density, $\mu$ is the magnetic permeability, and $\eta$ is magnetic diffusivity. With these scalings, the dimensionless governing equations can be written as
 \begin{linenomath*}
 \begin{equation}
  \frac{\partial \boldsymbol u}{\partial t}+ \boldsymbol u \cdot \boldsymbol \nabla\boldsymbol u+2 \hat{\boldsymbol z} \times \boldsymbol u = -\boldsymbol \nabla P+E \boldsymbol \nabla^{2} \boldsymbol u+ ({Ra_{T}^*}T + {Ra_{C}^*}C \boldsymbol) \frac{\boldsymbol r}{r_o}+ \frac {E}{Pm}[(\boldsymbol \nabla\times \boldsymbol B)\times \boldsymbol B],
  \label{eq21}
 \end{equation}
 \end{linenomath*}
 \begin{linenomath*}
 
 \begin{equation}
  \frac{\partial T}{\partial t}+\boldsymbol u \cdot \boldsymbol \nabla T =\frac{E}{Pr} \boldsymbol \nabla^{2} T,
  \label{eq22}
 \end{equation}
 
 \end{linenomath*}
 \begin{linenomath*}
 \begin{equation}
  \frac{\partial C}{\partial t}+\boldsymbol u \cdot \boldsymbol \nabla C =\frac{E}{Sc} \boldsymbol \nabla^{2} C,
  \label{eq23}
 \end{equation}
 \end{linenomath*}
 \begin{linenomath*}
 \begin{equation}
  \frac{\partial \boldsymbol B}{\partial t} =\boldsymbol \nabla\times (\boldsymbol u\times \boldsymbol B)+ \frac{E}{Pm}\boldsymbol \nabla^{2}\boldsymbol B,
  \label{eq24}
 \end{equation}
 \end{linenomath*}
 \begin{linenomath*}
 \begin{equation}
  \boldsymbol \nabla \cdot \boldsymbol u=0,
  \label{eq25}
 \end{equation}
 \end{linenomath*}
 \begin{linenomath*}
 \begin{equation}
  \boldsymbol \nabla \cdot \boldsymbol B=0,
  \label{eq26}
 \end{equation}
 \end{linenomath*}
 where $\boldsymbol{u}$ denotes the velocity field, $P$ the reduced pressure, $T$ the temperature, $C$ the composition, and $\boldsymbol{B}$ the magnetic field. The system is characterized by six dimensionless control parameters: the Ekman number $E$, the modified thermal Rayleigh number $Ra_{T}^*$, the modified compositional Rayleigh number $Ra_{C}^*$, the Prandtl number $Pr$, the Schmidt number $Sc$ and the magnetic Prandtl number $Pm$: 
 \begin{linenomath*}
 \begin{equation}
  E=\frac{\nu}{\Omega D^2}, Ra_{T}^*=\frac{\alpha g_{0}\Delta T}{\Omega^2 D},  Ra_{C}^*=\frac{\beta  g_{0}\Delta C}{\Omega^2 D},   Pr=\frac{\nu}{\kappa_{T}},  Sc=\frac{\nu}{\kappa_{C}},  Pm=\frac{\nu}{\eta},
  \label{eq27}
 \end{equation}
 \end{linenomath*}
 where $\nu$ is the kinematic viscosity, $\alpha$ is the thermal expansion coefficient and $\beta$ is the compositional expansion coefficient. The rotationally modified thermal or compositional Rayleigh number $Ra_{T/C}^*$ is commonly used in studies of rotating convection and corresponds to the squared convective Rossby number \citep{Gilman01011977}. 

 In addition to the control parameters introduced above, several additional dimensionless quantities are particularly important for FDDC. Among these, the Lewis number $Le$ mentioned in Section \ref{sec:intro}, is defined as the ratio of the Schmidt number to the Prandtl number,
 \begin{linenomath*}
 \begin{equation}
  Le=\frac{\kappa_{T}}{\kappa_{C}}=\frac {Sc}{Pr}.
  \label{eq28}
 \end{equation}
 \end{linenomath*}
 To quantify the strength of the TSS, we introduce the density ratio $R_{\rho}$ \citep{Stern1960}, defined as
  \begin{linenomath*}
 \begin{equation}
  R_{\rho}=\frac{\left | \alpha\Delta T \right | }{\beta \Delta C}=\frac{\left | Ra_{T}^* \right | }{Ra_{c}^*}.
  \label{eq29}
 \end{equation}
 \end{linenomath*}
 Here, the modified thermal Rayleigh number $Ra_T^*$ is negative, reflecting the stabilising effect of the TSS, and its absolute value is therefore used. In this formulation, the classical instability regime for FDDC corresponds to $1 \le  R_{\rho} < Le$ \citep{baines1969thermohaline}. Subsequent linear stability analyses demonstrated that this instability range remains valid in rotating planar systems \citep{Sengupta_2018}.
 
To further characterize the relative influence of TSS and rotation, we introduce the thermal buoyancy frequency, which quantifies the contribution of the thermal stratification alone. The square of the thermal buoyancy frequency normalized by the fluid angular velocity $\Omega$ is defined as
 \begin{linenomath*}
\begin{equation}
  \frac{N_T^2}{\Omega^2}(r) = Ra_{T}^*\frac{dT_c}{dr},
  \label{equa210}
\end{equation}
\end{linenomath*}
where $T_c$ denotes the conductive temperature profile adopted from \citet{gastine2015turbulent}, given by
\begin{linenomath*}
\begin{equation}
  T_{c}(r)=\frac{\eta}{(1-\eta)^2}\frac{1}{r}-\frac{\eta}{1-\eta},
  \label{equa211}
\end{equation}
\end{linenomath*}
where $\eta = r_i/r_o $ is the radius ratio. Although the conductive compositional profile has the same radial dependence, only the thermal contribution is retained in the definition of $N_T^2$, since the objective of this study is to quantify the influence of TSS. In the following, the value of $N_T^2/\Omega^2$ evaluated at the inner-core boundary is used as a representative measure of the thermal stratification strength.

To complete the model description, we now specify the boundary and initial conditions. The radius ratio is fixed at $\eta = r_i/r_o = 0.35$. Both boundaries are impermeable, electrically insulating, no-slip and maintained at constant temperatures and compositions. The initial velocity field is set to zero, while the initial temperature , composition, and magnetic fields are initialized with random perturbations. 

 To quantify the dynamical state of the system, we introduce several diagnostic quantities. The magnetic energy $E_b$ and kinetic energy $E_u$ are defined as
 \begin{linenomath*}
 \begin{equation}
  E_b=\frac{1}{2Pm/E} \int_{V} \boldsymbol B^2 \mathrm{d}V, \quad E_u=\frac{1}{2} \int_{V} \boldsymbol u^2 \mathrm{d}V,
  \label{eqebeu}
 \end{equation}
 \end{linenomath*}
 where $V$ is spherical shell of volume. Based on the adopted non-dimensionalisation, the Reynolds number can be expressed in terms of the kinetic energy as
 \begin{linenomath*}
 \begin{equation}
  Re=\frac{1}{E}\sqrt{\frac{2E_u}{V}}.
  \label{eqRe}
 \end{equation}
 \end{linenomath*}
 The magnetic Reynolds number and Rossby number are then obtained as
 \begin{linenomath*}
 \begin{equation}
  Rm=RePm, \quad Ro=ReE.
  \label{eqRm}
 \end{equation}
 \end{linenomath*}
 In addition, we define the local Rossby number
 \begin{linenomath*}
 \begin{equation}
  Ro_{\ell}=Ro\frac{D}{L_u},
  \label{eqRo}
 \end{equation}
 \end{linenomath*}
 where $L_u$ is the typical flow length scale and is determined from the time-averaged kinetic energy spectrum following \cite{christensen2006scaling}
\begin{linenomath*}
 \begin{equation}
L_u^{-1} = \overline{\left(\frac{D \sum\limits_{l = 1}^{l_{\max }} \sum\limits_{m = 0}^{l} l \mathcal{E}_{l}^{m}(t)}{\pi \sum\limits_{l = 1}^{l_{\max }} \sum\limits_{m = 0}^{l} \mathcal{E}_{l}^{m}(t)}\right)},
  \label{length}
 \end{equation}
 \end{linenomath*}
 where $\mathcal{E}_{l}^{m}$ is the dimensionless kinetic energy at a spherical harmonic degree $l$ and order $m$. Overbars ${\overline{\dots}}$ correspond to temporal averaging. The magnetic field
length scale $L_b$ is also determined following the same definition as that of the flow field length scale. 

The thermal and compositional buoyancy power $P_T$ and $P_C$ defined by
\begin{linenomath*}
 \begin{equation}
P_{T}(t) =V\left\langle{Ra_T^*} T(\boldsymbol{r}, t) \frac{r}{r_{o}} u_{r}(\boldsymbol{r}, t)\right\rangle_V, \quad P_{C}(t) =V\left\langle{Ra_C^*} C(\boldsymbol{r}, t) \frac{r}{r_{o}} u_{r}(\boldsymbol{r}, t)\right\rangle_V,
  \label{power}
 \end{equation}
 \end{linenomath*}
where $\langle \dots\rangle_V$ correspond to an average over a spherical shells.
 To quantify the strength of the magnetic field, we introduce the Elsasser number 
 \begin{linenomath*}
 \begin{equation}
  \Lambda =\frac{B_{rms}^2}{\rho \mu \eta \Omega},
  \label{eqElsasser}
 \end{equation}
 \end{linenomath*}
 where $B_{rms}$ is the dimensional root-mean-square (r.m.s.) magnetic.
 To characterize the helical properties of the flow, we calculate the relative helicity, with reference to \cite{schaeffer2017turbulent}, as
 \begin{linenomath*}
\begin{equation}
H(r,\theta)=\left\langle \frac{\boldsymbol{u}\cdot\boldsymbol{\omega}}{|\boldsymbol{u}|\,|\boldsymbol{\omega}|} \right\rangle_\phi ,
\label{eq:relative_helicity_avg}
\end{equation}
\end{linenomath*}
 where $\boldsymbol{u}$ and $\boldsymbol{\omega}=\nabla \times \boldsymbol{u} $ denote the velocity and vorticity fields, respectively. The $\langle \dots\rangle_{\phi}$ denotes the average over longitude $\phi$.

\subsection{Numerical method}
\label{sec:method}
We use the open-source code XSHELLS (\url{https://www.bitbucket.org/nschaeff/xshells/}) to solve the governing equations (\ref{eq21}-\ref{eq26}) subjected to the boundary conditions. The fluid is assumed to be incompressible, and the velocity $\boldsymbol u$ and magnetic $\boldsymbol B$ can be decomposed into toroidal and poloidal components:
 \begin{linenomath*}
 \begin{equation}
 \boldsymbol u=\boldsymbol \nabla \times(T\boldsymbol r)+\boldsymbol \nabla \times \boldsymbol \nabla \times (P\boldsymbol r),
 \label{eq220}
 \end{equation}
 \end{linenomath*}
 \begin{linenomath*}
 \begin{equation}
 \boldsymbol B=\boldsymbol \nabla \times(\mathcal{T} \boldsymbol r)+\boldsymbol \nabla \times \boldsymbol \nabla \times(\mathcal{P} \boldsymbol r).
 \label{eq210}
 \end{equation}
 \end{linenomath*}
 The toroidal $T$ and $\mathcal{T}$, poloidal $P$ and $\mathcal{P}$ scalar fields, as well as the temperature field $T$ and compositional field $C$, are expanded using spherical harmonic expansion on spherical surfaces. XSHELLS employs a second-order finite differences method in the radial direction, along with a pseudo-spectral spherical harmonic expansion. The spectral expansion is truncated up to spherical harmonics of degree $l_{max}$ and $Nr$ denotes the number of radial grid points. 
 To enhance computational efficiency, the code utilizes the SHTns library for fast spherical harmonic transformations \citep{schaeffer2013efficient}. 

\section{Results}
\label{sec:results}
\subsection{Model parameters}
\label{sec:Model parameters}
For FDDC dynamos, the system is governed by six dimensionless control parameters, leading to a large parameter space. Since the primary objective of this study is to investigate the influence of TSS on the dynamo process, we vary only the thermal Rayleigh number $Ra_T^*$ to control the strength of the TSS, while keeping the other five control parameters fixed. The Ekman number is set to $E = 3 \times 10^{-5}$, a value commonly used in numerical dynamo studies. The Prandtl number and the Schmidt number are chosen as $Pr = 0.1$ and $Sc = 10$, respectively, corresponding to a Lewis number $Le = 100$ and ensuring a strong contrast between thermal and compositional diffusivities. 
To resolve the compositional fluctuations associated with FDDC at such high Lewis numbers, we employ high spatial resolutions. The detailed numerical resolutions adopted for each simulation are provided in Table \ref{AppendixA} of the Appendix.
The magnetic Prandtl number is set to $Pm = 5$, which allows the system to sustain a strong-field dynamo regime in the absence of stable stratification. The compositional Rayleigh number is fixed at $Ra_C^* = 0.027$, corresponding to the geostrophic turbulence regime for high-Prandtl-number rotating convection identified by \citet{Fan_Wang_Lin_2024}. This regime is characterized by complex turbulent flow together with efficient heat transport while remaining under strong rotational constraint.

\begin{figure}
   \centering
   \includegraphics[width=0.9\linewidth]{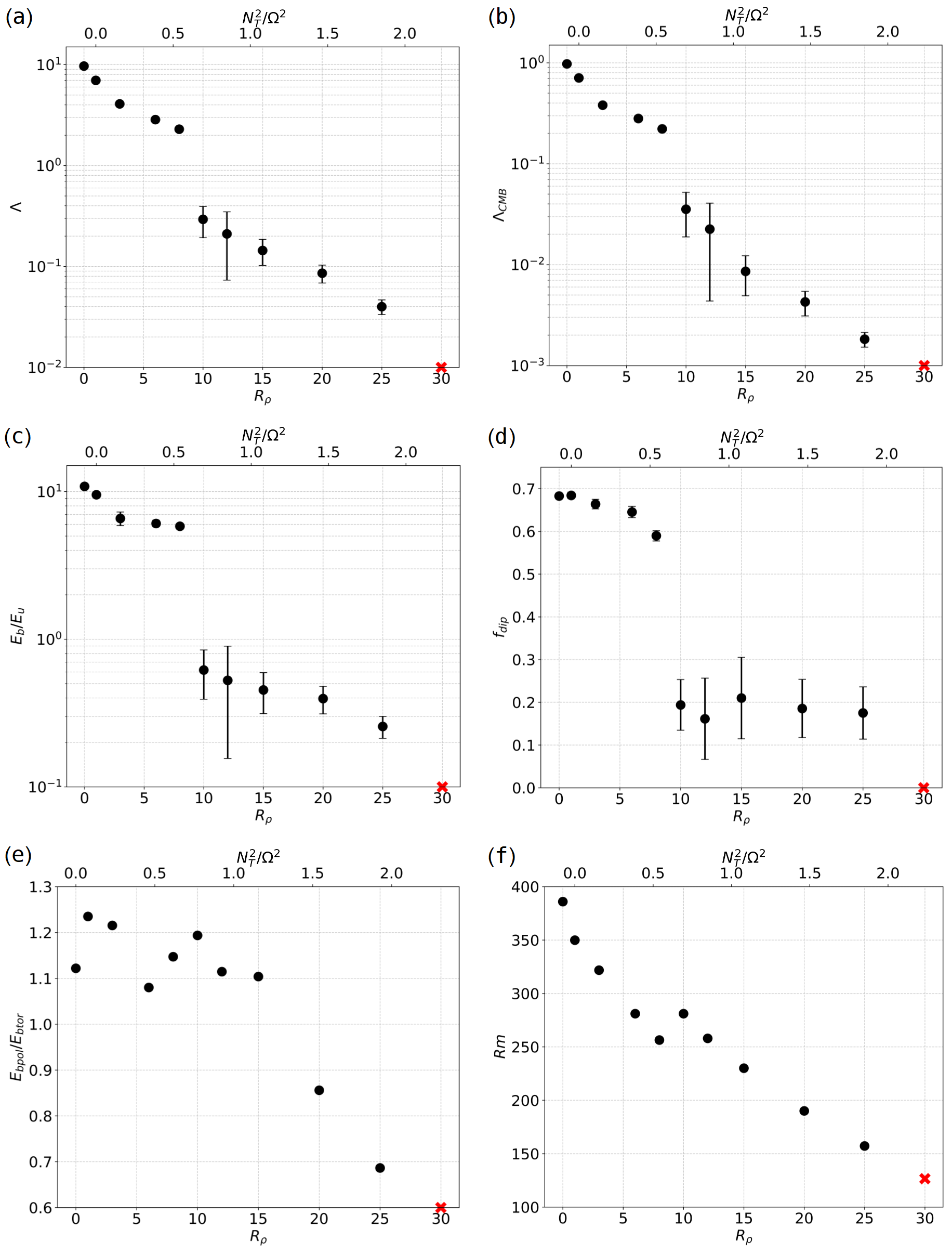}
   \caption{Variation of key dynamo diagnostics with strengthening thermally stable stratification. (a) Elsasser number; (b) Elsasser number evaluated at the core-mantle boundary; (c) ratio of magnetic to kinetic energy; (d) dipolarity, following \cite{christensen2006scaling}; (e) ratio of poloidal to toroidal magnetic energy; (f) magnetic Reynolds number. Red crosses denote cases without dynamo action and are plotted along the horizontal axis, indicating that the corresponding quantities are numerically zero.}
   \label{fig:1}
\end{figure}

\subsection{Dynamo simulations}
\label{sec:Dynamo simulations}

We adopt a strategy of increasing the strength of the TSS to mimic the effects of long-term planetary cooling on the dynamo behaviour. We first compute a reference dynamo without TSS until a statistically steady state is reached. The resulting fields are then used as initial conditions for simulations with weak TSS ($R_{\rho}=1$), allowing us to assess its impact on magnetic field morphology and intensity. Once a new statistically steady state is established, $R_{\rho}$ is further increased. This procedure is repeated iteratively until the magnetic energy exhibits sustained decay and approaches zero, indicating the cessation of dynamo action.


The evolution shown in Figure \ref{fig:1} indicates that, as TSS strengthens, the dynamo does not undergo a smooth, gradual decay. Instead, it exhibits a transition between distinct dynamical regimes over a finite range of TSS strength. As shown in Figure \ref{fig:1}(a,b), increasing $R_{\rho}$ progressively suppresses convective motions, leading to a gradual decrease in both the magnetic energy within the spherical shell and that at the CMB. However, once the TSS exceeds a critical threshold ($R_{\rho}\approx 10$), the system undergoes a qualitative change, marked by a rapid collapse of magnetic energy.  This abrupt reduction signals a transition in the dynamo regime. As illustrated in Figure \ref{fig:1}(c), the system shifts from a strong-field state to a weak-field state in which magnetic feedback on the flow is substantially reduced. This transition is also reflected in the magnetic field morphology: the field evolves from a dipole-dominated configuration to a multipolar structure, as evidenced by the sharp decrease in dipolarity shown in Figure \ref{fig:1}(d). Furthermore, the evolution of the ratio of poloidal to toroidal magnetic energy (Figure \ref{fig:1}(e)) differs from that of the total magnetic energy. Unlike the abrupt reduction in magnetic energy at $R_{\rho}=10$, the poloidal-to-toroidal magnetic energy ratio does not exhibit a significant change at this density ratio. Instead, a pronounced decrease is observed at $R_{\rho}=20$. By comparison, the flow responds immediately to the magnetic energy collapse. The rapid decline in magnetic energy is accompanied by a temporary enhancement of flow intensity during the weak-field phase, as reflected by the modest increase in the magnetic Reynolds number (Figure \ref{fig:1}(f)). As TSS continues to strengthen, convective motions are suppressed, leading to a subsequent decrease in the magnetic Reynolds number and ultimately to the cessation of dynamo action.

\begin{figure}[t]
   \centering
   \includegraphics[width=0.98\linewidth]{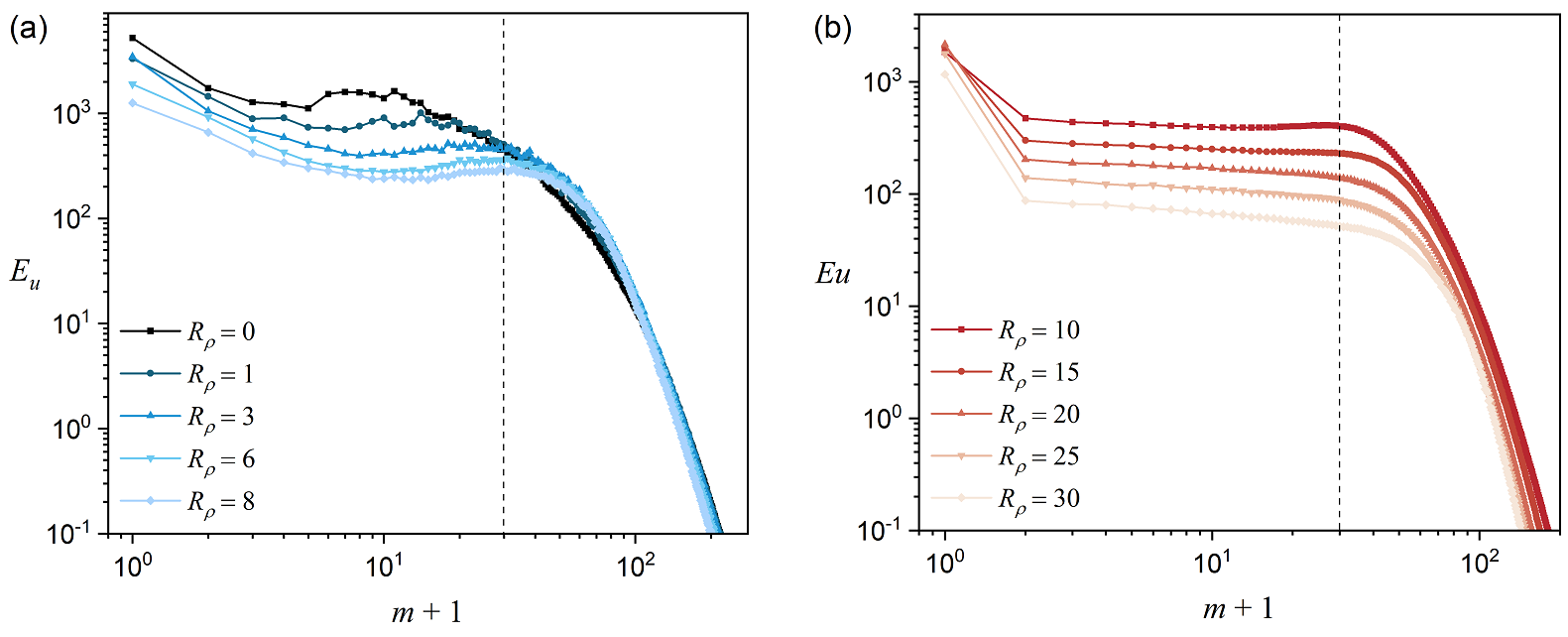}
   \caption{Azimuthal wavenumber kinetic energy spectra for different strengths of thermally stable stratification: (a) strong-field dynamo cases; (b) weak-field dynamo cases.}
    \label{fig:2}
\end{figure}

\begin{figure}[t]
   \includegraphics[width=\linewidth]{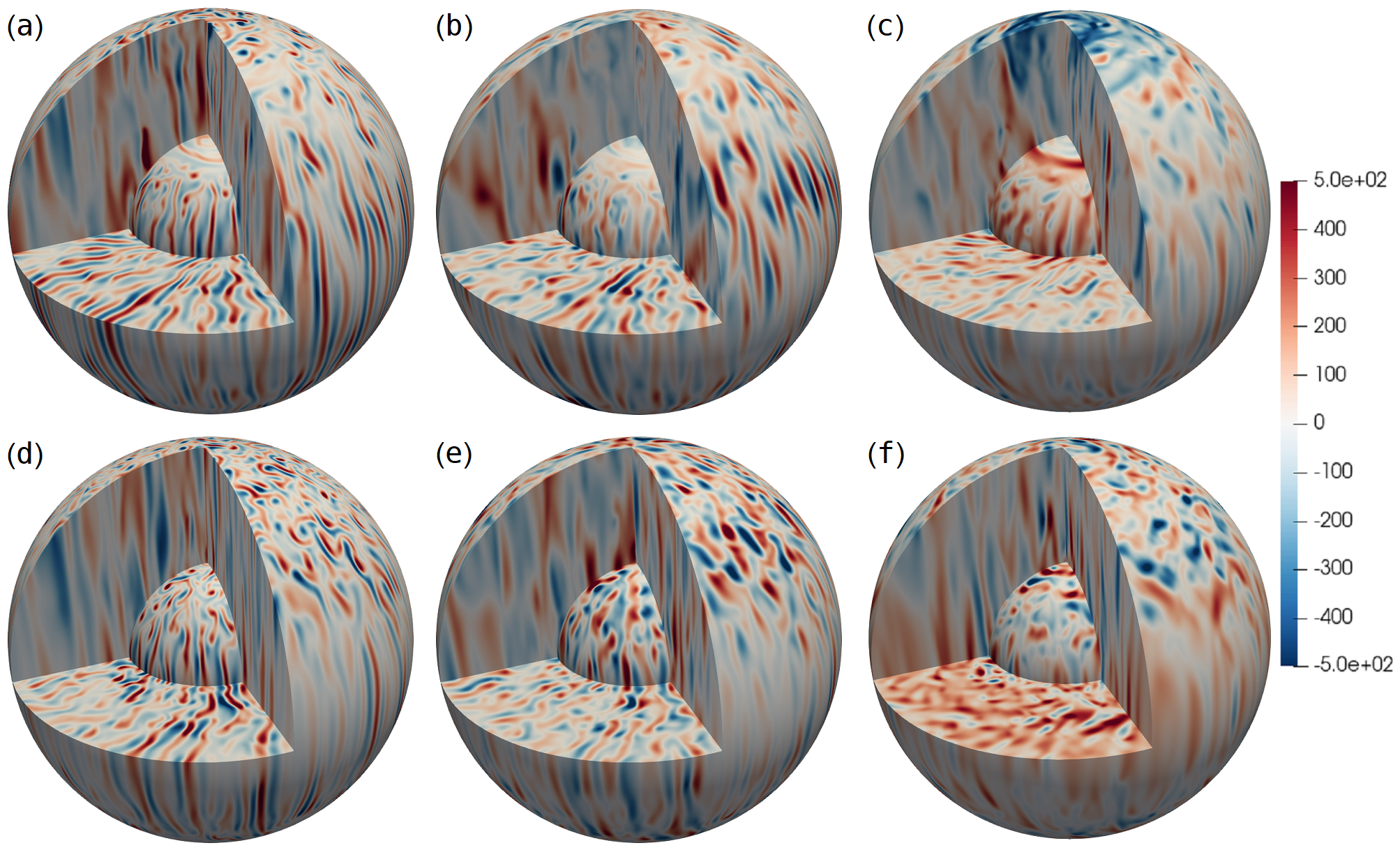}
   \caption{Comparison of three-dimensional snapshots of the velocity components for different strengths of thermally stable stratification. The inner (outer) surface corresponds to a spherical surface at radius $r_i+0.1D$ $(r_o-0.1D)$.
Panels (a)-(c) show the radial ($u_r$), latitudinal ($u_\theta$), and azimuthal ($u_\phi$) velocity components for the case with $R_{\rho}=8$, while panels (d)-(f) show the corresponding components for $R_{\rho}=10$. The colour bar represents the velocity amplitude, expressed in terms of the magnetic Reynolds number.}
   \label{fig:3}
\end{figure}

 Further analysis of the kinetic energy spectra (see Figure \ref{fig:2}) reveals that the transition from the strong-field to the weak-field regime involves not only a reduction in kinetic energy, but more importantly a reorganisation of flow structure. 
 In the strong-field regime, the kinetic energy decreases smoothly with increasing azimuthal wavenumber, indicating a continuous distribution of energy from large to small scales. By contrast, the weak-field regime is distinguished by a pronounced axisymmetric ($m=0$) component, reflecting the enhanced contribution of zonal flow. Beyond the axisymmetric mode, the kinetic energy drops sharply and then remains nearly constant over the range $m=1-30$. These contrasting spectral characteristics indicate that the flow organization differs fundamentally between the strong- and weak-field dynamo regimes.

\subsubsection{Transition from  strong-field to  weak-field dynamo}

To further elucidate how changes in flow morphology are linked to the abrupt collapse of magnetic energy, we examine the three components of the three-dimensional velocity field for two representative cases with $R_{\rho} = 8$ and $R_{\rho} = 10$. As shown in Figure \ref{fig:3}, the two cases exhibit only minor differences in $u_r$ and $u_{\theta}$ components, while the primary differences arise in $u_{\phi}$. In the $R_{\rho} = 8$ case, pronounced retrograde zonal flow is observed at high latitudes, consistent with the strong-field dynamo regime reported by \cite{schaeffer2017turbulent}. In contrast, when TSS strengthens to $R_{\rho} = 10$, the retrograde zonal flow at high latitudes disappears, and a pronounced prograde zonal flow develops in the equatorial and mid-latitude regions. A similar emergence of equatorial prograde zonal flow with strengthening TSS has been reported by \cite{monville2019rotating} and \cite{GRAY2026107570}. Despite these changes in flow morphology, the overall flow remains strongly constrained by rotation. The flow retains clear signatures of columnar organization, and the local Rossby numbers listed in Table \ref{AppendixA} of the Appendix further indicate that both cases remain within the strongly rotationally constrained regime.

\begin{figure}
\centering
   \includegraphics[width=0.8\linewidth]{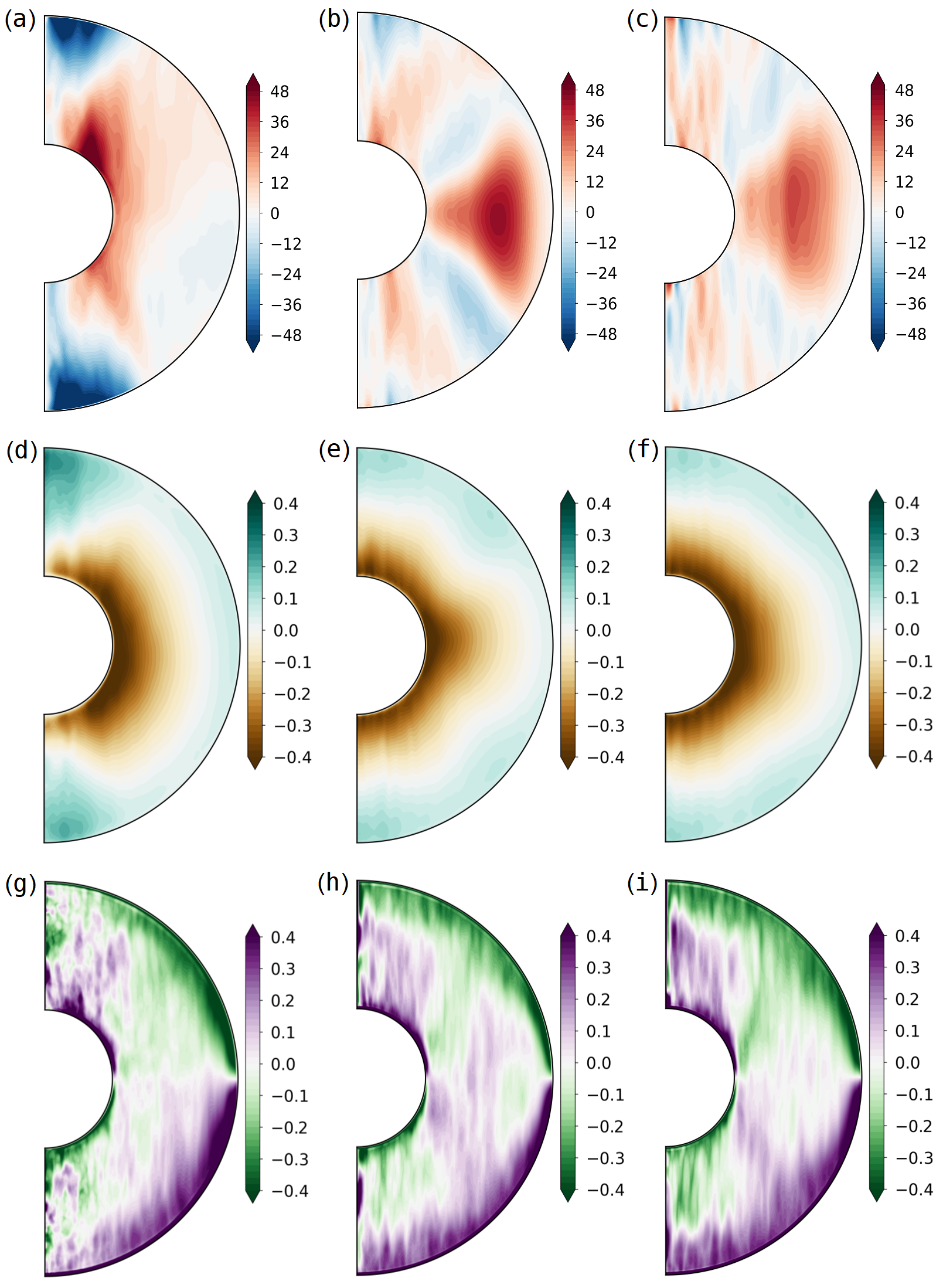}
   \caption{Meridional-plane distributions of the time- and longitude-averaged zonal flow, compositional perturbation, and relative helicity. Panels (a)--(c) show the zonal flow for the dynamo case with $R_{\rho}=8$, the dynamo case with $R_{\rho}=10$, and the hydrodynamic case with $R_{\rho}=8$, respectively. Panels (d)--(f) show the corresponding compositional perturbation fields, normalized by the background compositional field at the outer boundary. Panels (g)--(i) show the corresponding relative helicity fields. The color bar in (a)--(c) represents the Reynolds number.}
   \label{fig:4}
\end{figure}

Because the instantaneous $u_\phi$ fields differ markedly between the two cases, we further apply time and longitude averaging to $u_\phi$ in order to provide a more quantitative and objective comparison of the zonal flow structures. As shown in Figure \ref{fig:4}(a) and \ref{fig:4}(b), the time- and longitude-averaged $u_\phi$ exhibits pronounced differences in spatial organization between the two cases. In the $R_{\rho}=8$ case (Figure \ref{fig:4}(a)), the zonal flow is primarily confined within the tangent cylinder. In contrast, in the $R_{\rho}=10$ case (Figure \ref{fig:4}(b)), the dominant zonal flow shifts to regions outside the tangent cylinder. This indicates a fundamental reorganisation of large-scale flow dynamics across the transition.
\begin{figure}[t]
   \includegraphics[width=\linewidth]{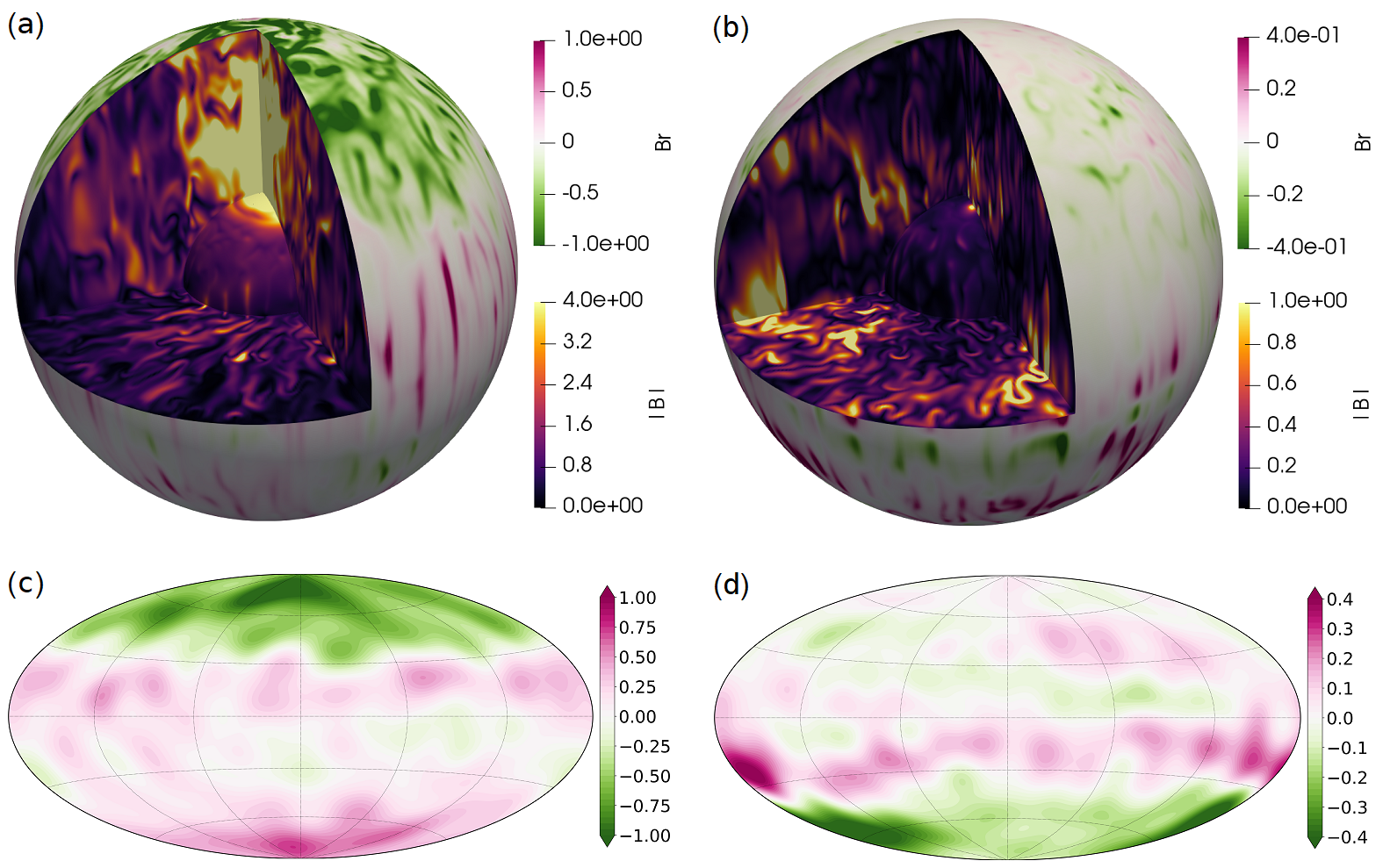}
   \caption{Comparison of magnetic field structures for the cases with $R_{\rho}=8$ and $R_{\rho}=10$. Panels (a, b) show three-dimensional visualizations of the magnetic field for $R_{\rho}=8$ and $R_{\rho}=10$, respectively. The outer surface represents the radial magnetic field, while the interior shows the magnetic field intensity. Panels (c, d) present the radial magnetic field at the core-mantle boundary (CMB) in Aitoff projection, truncated at spherical harmonic degree $l=13$, for $R_{\rho}=8$ and $R_{\rho}=10$, respectively. The magnetic field is in (non‐dimensional) Elsasser units (scaled by $\sqrt{\rho \mu \eta \Omega}$).}
   \label{fig:5}
\end{figure}
However, based on the results above alone, it is difficult to determine whether changes in the zonal flow structure trigger the abrupt drop in dynamo magnetic energy, or whether the pronounced decay of magnetic energy weakens the magnetic suppression of the flow, thereby leading to an adjustment of the flow structure. To clarify this causal relationship, we additionally performed a purely hydrodynamic simulation at $R_{\rho}=8$ for comparison (see Figure \ref{fig:4}(c)). The results show that, even in the absence of magnetic feedback, changes in the zonal flow structure already emerge at $R_{\rho}=8$. This indicates that the dynamo at $R_{\rho}=8$ is already close to the threshold for the abrupt collapse of magnetic energy. When $R_{\rho}$ is further increased to 10, the continued reorganization of the zonal flow exceeds this threshold, rendering the strong-field dynamo unsustainable. We also compare the compositional perturbation fields of the three cases, as shown in Figures \ref{fig:4}(d-f). The strong-field dynamo exhibits stronger compositional perturbations at high latitudes, whereas the weak-field and hydrodynamic cases show a comparatively more uniform perturbation distribution. In addition, the relative helicity fields shown in Figures \ref{fig:4}(g-i) indicate that the reorganization of the flow morphology is not accompanied by a significant change in helicity magnitude.
\begin{figure}[t]
   \includegraphics[width=\linewidth]{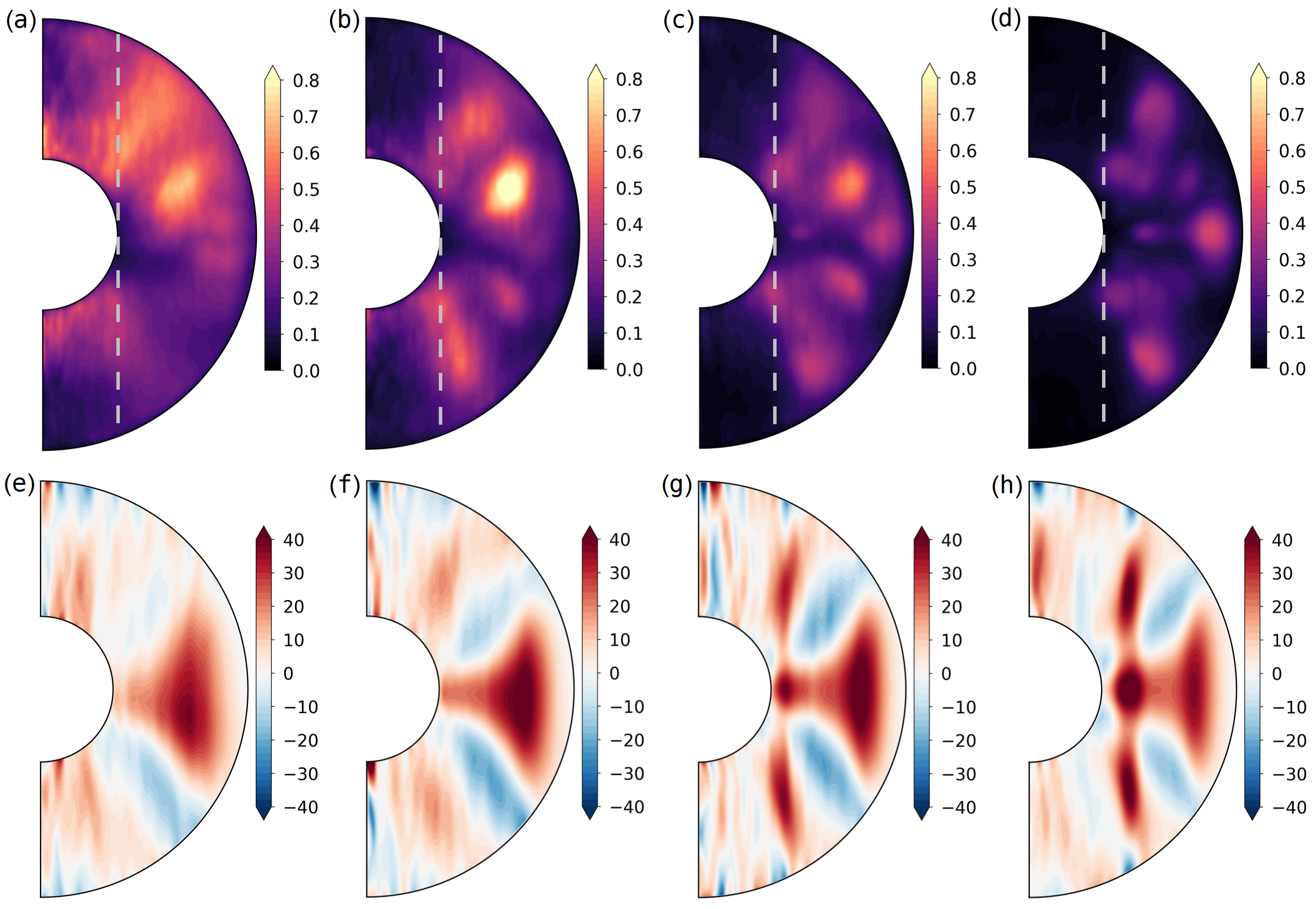}
   \caption{Time- and longitude-averaged magnetic field intensity and zonal flow for weak-field dynamos under different strengths of thermally stable stratification. Panels (a-d) show the magnetic field intensity, and panels (e-h) show $u_{\phi}$. (a, e) $R_{\rho}=12$; (b, f) $R_{\rho}=15$; (c, g) $R_{\rho}=20$; (d, h) $R_{\rho}=25$. The magnetic field is in (non-dimensional) Elsasser units (scaled by $\sqrt{\rho \mu \eta \Omega}$).}
   \label{fig:6}
\end{figure}

The pronounced changes in zonal flow not only modulate the strength and dipolarity of the magnetic field, but also exert a profound influence on its internal organization. The spatial distribution of the internal magnetic field intensity in Figure \ref{fig:5}(a) is broadly consistent with the strong-field dynamo characteristics reported by \cite{schaeffer2017turbulent}, with the magnetic field being stronger inside the tangent cylinder than outside, resulting in a largely inhomogeneous spatial distribution. In contrast, in the weak-field dynamo case with $R_{\rho}=10$ (Figure \ref{fig:5}(b)), the magnetic field is no longer stronger inside the tangent cylinder than outside and exhibits more pronounced small-scale magnetic structures. The corresponding magnetic field length scales are listed in Table \ref{AppendixA} of the Appendix. Moreover, the radial magnetic field distribution at the CMB corresponds closely to the dipolarity shown in Figure \ref{fig:1}(d), clearly indicating a transition of the magnetic field from a dipole-dominated to a multipolar-dominated state. 

\subsubsection{Pathway to dynamo cessation}

Once the dynamo system enters a weak-field regime dominated by multipolar magnetic fields, magnetic-field-related diagnostics exhibit markedly enhanced temporal variability. To further characterize the evolution of magnetic field morphology under weak-field conditions,  Figure \ref{fig:6}(a-d) presents the spatial distribution of magnetic field intensity for a range of TSS strengths. At $R_{\rho} = 12$, the magnetic field distribution appears relatively disordered and exhibits a certain degree of north-south hemispheric asymmetry. As $R_{\rho}$ increases to 15, the field becomes more organized and displays a clear equatorial symmetry, while the magnetic field intensity within the tangent cylinder is weakened. With a further increase to $R_{\rho}=20$, the overall magnetic field intensity continues to decrease. Finally, at $R_{\rho}=25$, the equatorial symmetry is further strengthened, and the tangent cylinder becomes nearly devoid of appreciable magnetic field intensity.
\begin{figure}[t]
   \includegraphics[width=\linewidth]{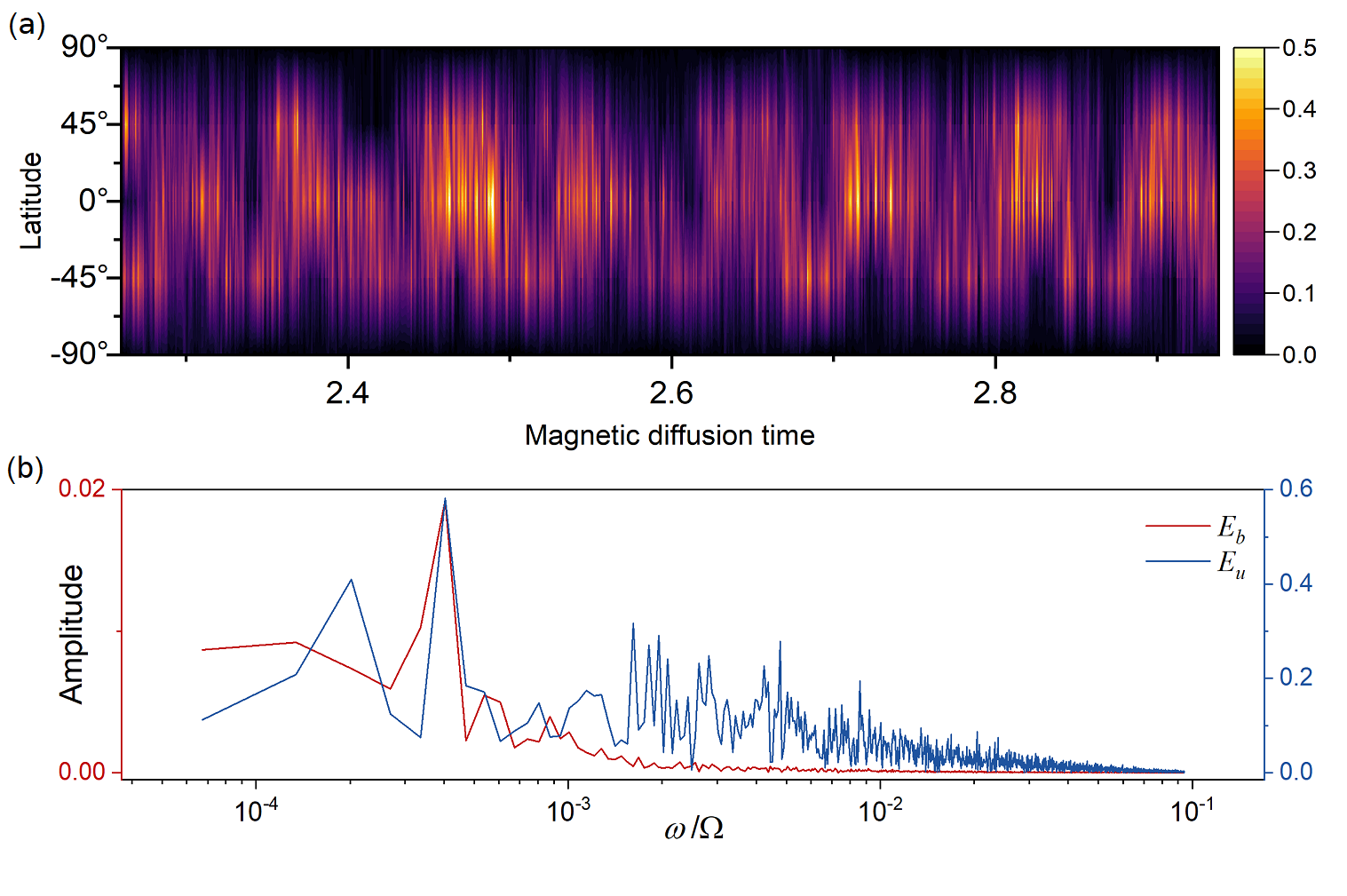}
   \caption{Temporal evolution and frequency characteristics of the magnetic field and flow for the $R_{\rho}=25$ dynamo case. (a) Time-latitude diagram of magnetic field intensity. (b) Fast Fourier transform spectra of the temporal variations of the kinetic and magnetic energies.}
   \label{fig:7}
\end{figure}

To elucidate the dynamical origin of this trend, the corresponding zonal flow distributions are shown in Figure \ref{fig:6}(e-h). As $R_{\rho}$ increases, the global magnetic Reynolds number gradually decreases (Figure \ref{fig:1}(f)), whereas the amplitude of the zonal flow does not diminish and instead shows an increasing tendency outside the tangent cylinder. Meanwhile, the ratio of magnetic to kinetic energy continuously decreases (Figure \ref{fig:1}(c)), indicating a progressive weakening of the magnetic feedback on the flow. As a result, the flow increasingly governs the magnetic field structure, leading to more equatorially symmetric field that is predominantly concentrated outside the tangent cylinder.

Since $R_{\rho}=25$ is close to the critical condition for dynamo cessation, we perform a more detailed analysis of the temporal evolution of the magnetic field in this case. As shown in Figure \ref{fig:7}(a), the magnetic field intensity is relatively strong at mid- to low latitudes, whereas it is weaker near the poles. In addition, the temporal evolution of the magnetic field exhibits a pronounced periodic feature. To further characterize this temporal variability, we perform fast Fourier transform (FFT) analyses of the volume-integrated magnetic and kinetic energies (Figure \ref{fig:7}(b)). The magnetic energy spectrum exhibits a clear dominant frequency, indicating a coherent oscillation of the magnetic field. Interestingly, the kinetic energy spectrum also shows a peak at the same frequency, suggesting that the magnetic field oscillation is closely associated with the underlying flow dynamics. Compared with the magnetic energy spectrum, the kinetic energy spectrum contains additional higher-frequency components, reflecting the multi-scale nature of the convective flow, where various convective motions coexist with the dominant oscillatory mode.

\section{Discussion}
\label{sec:discussion}
Based on the numerical results presented above, the evolution of the dynamo under strengthened TSS can be interpreted as a competition between Lorentz-force control and stratification-induced flow reorganization. Despite the pronounced differences in flow morphology across the strong- to weak-field transition (i.e., between the $R_{\rho}=8$ and $R_{\rho}=10$ cases), both flows remain under strong rotational constraint, and the relative helicity does not exhibit significant differences between the two dynamo states. We therefore attribute the dynamo transition primarily to the stratification-induced reorganization of the large-scale flow, rather than to changes in rotational constraint or helical properties. This reorganization is mainly associated with the emergence and strengthening of prograde zonal flow, which can be interpreted in terms of a thermocompositional wind balance, in which latitudinal variations of the axisymmetric density perturbations are balanced by axial variations of the zonal velocity \citep{GRAY2026107570}. 

During the early stages of planetary cooling, although a TSS has already developed, the dynamo system can remain in a strong-field state. In this regime, the Lorentz force imposes strong dynamical constraints on the flow, effectively limiting the development of large-scale circulations and suppressing flow reorganisation associated with TSS. Consequently, even as convective driving is weakened, the magnetic field can still maintain a relatively coherent and organised spatial structure.
However, once the strength of TSS exceeds a critical threshold, the Lorentz force is no longer sufficient to counteract the flow reorganization associated with the strengthened prograde zonal flow. The system then undergoes a rapid transition from a strong-field to a weak-field dynamo state, accompanied by an abrupt collapse of magnetic energy. With further strengthening of the TSS, the gradual emergence and intensification of prograde equatorial zonal flow reflect a continuous reorganization of the large-scale circulation. Although this flow reorganization is initially moderated by strong Lorentz-force feedback, the further strengthening of the clockwise zonal flow eventually alters the large-scale circulation beyond the ability of Lorentz forces to maintain the strong-field state, leading to the transition toward the weak-field dynamo regime.

Once the system enters the weak-field dynamo regime, the previously stable, dipole-dominated magnetic configuration is disrupted. The magnetic field becomes more multipolar and exhibits enhanced temporal variability. The time-averaged magnetic field distribution indicates that, with increasing strength of TSS, the magnetic field within the tangent cylinder gradually weakens and eventually nearly vanishes, while the field outside the tangent cylinder becomes increasingly equatorially symmetric. In addition, the temporal evolution of the magnetic field exhibits clear oscillatory behaviour. In the weak-field dynamo regime, where the magnetic energy is much smaller than the kinetic energy, the magnetic field evolution is expected to be primarily influenced by the flow rather than by magnetic feedback. This interpretation is supported by the FFT analysis, which shows that the dominant frequency in the magnetic energy spectrum coincides with that in the kinetic energy spectrum. Therefore, the low-frequency magnetic variation likely reflects the underlying hydrodynamic oscillation.

However, the physical nature of this oscillatory behaviour cannot be identified from the present analysis. Such an identification would require resolving the spatial propagation characteristics of the dominant mode. In our simulations, the convective flow contains numerous higher-frequency fluctuations superimposed on the dominant low-frequency oscillation. These fluctuations obscure the propagation characteristics of the dominant mode, making it difficult to identify its wave properties through direct visualization alone. Furthermore, we have not identified a clear correspondence between this characteristic frequency and any specific dynamical timescale in the system. Therefore, we refrain from attributing this oscillatory behaviour to a specific wave type. Nevertheless, this oscillatory behaviour appears only as an additional feature of the weak-field regime and does not alter the overall trend that increasing TSS weakens the dynamo.

The overall weakening of the dynamo with increasing TSS suggests that, as the system approaches its cessation threshold, the interaction between convective motions and TSS alters the magnetic field structure. With further strengthening of the TSS, convective intensity continues to decline and the magnetic Reynolds number correspondingly decreases, eventually falling below the critical value required to sustain self-excited dynamo action. The resulting cessation demonstrates that even if compositional convection persists, for example due to inner-core crystallisation and the associated release of light elements, dynamo action may still be suppressed when TSS becomes strong. This highlights the critical role of TSS in regulating the long-term evolution of planetary dynamos, showing that the presence of compositional buoyancy alone does not necessarily guarantee the sustained generation of a magnetic field, as its effectiveness is strongly constrained by the thermal structure of the core. 

In the context of Mars, earlier studies have generally favoured the absence of a solid inner core \citep[e.g.][]{stevenson2001mars,stahler2021seismic}, implying that compositional convection associated with inner-core growth would not contribute to dynamo activity. However, recent seismological analyses \citep{bi2025seismic} suggest that Mars may possess a solid inner core with a radius of approximately 600 km. This raises an apparent inconsistency, as compositional convection driven by inner-core crystallisation could be active, yet Mars does not exhibit a present-day global intrinsic magnetic field \citep{Mittelholz}. In light of our results, TSS in the Martian outer core may provide a plausible explanation for this discrepancy. Even in the presence of compositional buoyancy, strong TSS could inhibit the ability of convective motions to sustain magnetic field generation, potentially leading to the cessation of the Martian dynamo.

\section{Conclusions}
\label{conclusions}
Using numerical simulations, we have examined how strengthened TSS influences rotating convective dynamos. Our results show that the dynamo evolution is controlled by a competition between Lorentz-force regulation and stratification-induced flow reorganization, leading to a sharp transition between distinct dynamo regimes. When TSS is weak, the dynamo can remain in a strong-field state despite reduced convective driving, with a stable, dipole-dominated magnetic field. As TSS intensifies, systematic changes in flow morphology emerge, including the development of prograde equatorial zonal flow. Once a critical threshold is exceeded, Lorentz force can no longer prevent flow reorganization, triggering an abrupt transition to a weak-field dynamo accompanied by a rapid collapse of magnetic energy. In the weak-field regime, the magnetic field becomes multipolar and exhibits enhanced temporal variability. With further strengthening of TSS, the field becomes increasingly equatorially symmetric and displays pronounced oscillatory behaviour. Ultimately, dynamo action ceases as convective intensity declines and the magnetic Reynolds number falls below the critical threshold, even in the presence of ongoing compositional convection. 

Our results also suggest that dynamo action in FDDC is sensitive to the governing parameters. This may explain the discrepancy with the decaying dynamos reported by \citet{mather2021regimes}. Compared with their study, the present simulations employ a lower Ekman number, a higher compositional Rayleigh number, and a larger Lewis number ($Le=100$), resulting in more vigorous convection under stronger rotational constraint. Moreover, the larger Lewis number allows the density ratio to span a much broader parameter range ($1\le R_{\rho}<Le$), reducing the sensitivity of the dynamo behaviour to variations in $R_{\rho}$. These differences suggest that the boundary between sustaining and decaying dynamos in FDDC is non-trivial and deserves further investigation.

Across this broad parameter range, our results further demonstrate that TSS acts as a key regulator of planetary dynamos. Even when compositional convection remains active, sufficiently strong TSS can reorganize the flow, suppress magnetic field generation, and eventually terminate dynamo action. These findings highlight the important role of the core thermal structure in controlling the long-term evolution and eventual cessation of planetary magnetic fields.

\section*{Acknowledgements}
This study was supported by the National
Key R\&D Program of China
(2022YFF0503200) and the National
Natural Science Foundation of China
(Grant Nos. 42350002, 12250012, 12595301). Numerical calculations were performed on the Taiyi cluster supported by the Center for Computational Science
and Engineering of Southern University of Science and Technology.

\appendix
\section{List of numerical simulations}

Table \ref{tabA1} lists the additional diagnostic quantities and numerical information for all numerical simulations presented in this study.  We also created a more comprehensive Excel spreadsheet containing all control parameters, diagnostic parameters, and numerical setup parameters, which is made publically available on Zenodo (\url{https://zenodo.org/records/21602262}).

\begin{table}[H]
 \caption{Diagnostics and numerical resolutions from simulations. The running time ($t_{run}$) and averaging time ($t_{avg}$) are given in units of the magnetic diffusion time, $\tau_{\eta}$. $N_r$ denotes the number of radial grid points, and $l_{\max}$ is the maximum spherical harmonic degree.}
 \label{tabA1}
 \centering
 \resizebox{\linewidth}{!}{
 \begin{tabular}{l cccccccc}
 \hline
  ~~Case&$L_u$&$L_b$&$P_T(\times10^7)$&$P_C(\times10^8)$&$Ro_{\ell}$&$t_{run}$&$t_{avg}$&$N_r\times l_{max}$\\
 \hline
   $R_{\rho}=0$ & 0.112& 0.153& 0    & 3.35& 0.021&0.065& 0.035& $400\times400$\\
   $R_{\rho}=1$ & 0.094& 0.153& -4.74& 2.85& 0.022&0.091& 0.043& $400\times400$\\
   $R_{\rho}=3$ & 0.081& 0.151& -7.41& 2.32& 0.024&0.169& 0.078& $380\times380$\\
   $R_{\rho}=6$ & 0.074& 0.158& -8.44& 1.85& 0.023&0.155& 0.072& $360\times360$\\
   $R_{\rho}=8$ & 0.069& 0.159& -8.57& 1.61& 0.020&0.303& 0.186& $320\times320$\\
   $R_{\rho}=10$& 0.070& 0.096& -9.17& 1.56& 0.024&2.317& 1.471& $280\times280$\\
   $R_{\rho}=12$& 0.069& 0.094& -8.43& 1.34& 0.022&2.666& 2.018& $280\times280$\\
   $R_{\rho}=15$& 0.070& 0.098& -7.12& 1.06& 0.020&1.867& 0.921& $280\times280$\\
   $R_{\rho}=20$& 0.074& 0.114& -5.21& 0.69& 0.015&2.446& 0.634& $280\times280$\\
   $R_{\rho}=25$& 0.076& 0.125& -3.88& 0.45& 0.012&2.051& 0.860& $256\times256$\\
 \hline
 \end{tabular}
 }
 \label{AppendixA}
 \end{table}



\begin{thebibliography}{46}
\expandafter\ifx\csname natexlab\endcsname\relax\def\natexlab#1{#1}\fi
\providecommand{\url}[1]{\texttt{#1}}
\providecommand{\href}[2]{#2}
\providecommand{\path}[1]{#1}
\providecommand{\DOIprefix}{doi:}
\providecommand{\ArXivprefix}{arXiv:}
\providecommand{\URLprefix}{URL: }
\providecommand{\Pubmedprefix}{pmid:}
\providecommand{\doi}[1]{\href{http://dx.doi.org/#1}{\path{#1}}}
\providecommand{\Pubmed}[1]{\href{pmid:#1}{\path{#1}}}
\providecommand{\bibinfo}[2]{#2}
\ifx\xfnm\relax \def\xfnm[#1]{\unskip,\space#1}\fi
\bibitem[{Baines and Gill(1969)}]{baines1969thermohaline}
\bibinfo{author}{Baines, P.}, \bibinfo{author}{Gill, A.}, \bibinfo{year}{1969}.
\newblock \bibinfo{title}{On thermohaline convection with linear gradients}.
\newblock \bibinfo{journal}{Journal of Fluid Mechanics} \bibinfo{volume}{37}, \bibinfo{pages}{289--306}.
\bibitem[{Bi et~al.(2025)Bi, Sun, Sun, Mao, Dai and Hemingway}]{bi2025seismic}
\bibinfo{author}{Bi, H.}, \bibinfo{author}{Sun, D.}, \bibinfo{author}{Sun, N.}, \bibinfo{author}{Mao, Z.}, \bibinfo{author}{Dai, M.}, \bibinfo{author}{Hemingway, D.}, \bibinfo{year}{2025}.
\newblock \bibinfo{title}{Seismic detection of a 600-km solid inner core in mars}.
\newblock \bibinfo{journal}{Nature} \bibinfo{volume}{645}, \bibinfo{pages}{67--72}.
\bibitem[{Bouffard et~al.(2017)Bouffard, Labrosse, Choblet, Fournier, Aubert and Tackley}]{BOUFFARD2017552}
\bibinfo{author}{Bouffard, M.}, \bibinfo{author}{Labrosse, S.}, \bibinfo{author}{Choblet, G.}, \bibinfo{author}{Fournier, A.}, \bibinfo{author}{Aubert, J.}, \bibinfo{author}{Tackley, P.J.}, \bibinfo{year}{2017}.
\newblock \bibinfo{title}{A particle-in-cell method for studying double-diffusive convection in the liquid layers of planetary interiors}.
\newblock \bibinfo{journal}{Journal of Computational Physics} \bibinfo{volume}{346}, \bibinfo{pages}{552--571}.
\bibitem[{Braginsky and Roberts(1995)}]{braginsky1995equations}
\bibinfo{author}{Braginsky, S.I.}, \bibinfo{author}{Roberts, P.H.}, \bibinfo{year}{1995}.
\newblock \bibinfo{title}{Equations governing convection in earth's core and the geodynamo}.
\newblock \bibinfo{journal}{Geophysical \& Astrophysical Fluid Dynamics} \bibinfo{volume}{79}, \bibinfo{pages}{1--97}.
\bibitem[{Breuer et~al.(2010)Breuer, Manglik, Wicht, Trümper, Harder and Hansen}]{breuer2010thermochemically}
\bibinfo{author}{Breuer, M.}, \bibinfo{author}{Manglik, A.}, \bibinfo{author}{Wicht, J.}, \bibinfo{author}{Trümper, T.}, \bibinfo{author}{Harder, H.}, \bibinfo{author}{Hansen, U.}, \bibinfo{year}{2010}.
\newblock \bibinfo{title}{{Thermochemically driven convection in a rotating spherical shell}}.
\newblock \bibinfo{journal}{Geophysical Journal International} \bibinfo{volume}{183}, \bibinfo{pages}{150--162}.
\bibitem[{Christensen and Aubert(2006)}]{christensen2006scaling}
\bibinfo{author}{Christensen, U.R.}, \bibinfo{author}{Aubert, J.}, \bibinfo{year}{2006}.
\newblock \bibinfo{title}{Scaling properties of convection-driven dynamos in rotating spherical shells and application to planetary magnetic fields}.
\newblock \bibinfo{journal}{Geophysical Journal International} \bibinfo{volume}{166}, \bibinfo{pages}{97--114}.
\bibitem[{Driscoll and Bercovici(2014)}]{DRISCOLL201436}
\bibinfo{author}{Driscoll, P.}, \bibinfo{author}{Bercovici, D.}, \bibinfo{year}{2014}.
\newblock \bibinfo{title}{On the thermal and magnetic histories of earth and venus: Influences of melting, radioactivity, and conductivity}.
\newblock \bibinfo{journal}{Physics of the Earth and Planetary Interiors} \bibinfo{volume}{236}, \bibinfo{pages}{36--51}.
\bibitem[{Fan and Lin(2025)}]{fan2025}
\bibinfo{author}{Fan, W.}, \bibinfo{author}{Lin, Y.}, \bibinfo{year}{2025}.
\newblock \bibinfo{title}{Dynamos driven by top-heavy double-diffusive convection in the strong-field regime}.
\newblock \bibinfo{journal}{Journal of Geophysical Research: Planets} \bibinfo{volume}{130}, \bibinfo{pages}{e2025JE008969}.
\bibitem[{Fan et~al.(2024)Fan, Wang and Lin}]{Fan_Wang_Lin_2024}
\bibinfo{author}{Fan, W.}, \bibinfo{author}{Wang, Q.}, \bibinfo{author}{Lin, Y.}, \bibinfo{year}{2024}.
\newblock \bibinfo{title}{Scaling behaviour of rotating convection in a spherical shell with different prandtl numbers}.
\newblock \bibinfo{journal}{Journal of Fluid Mechanics} \bibinfo{volume}{998}, \bibinfo{pages}{A20}.
\bibitem[{Gastine et~al.(2020)Gastine, Aubert and Fournier}]{Gastinestable}
\bibinfo{author}{Gastine, T.}, \bibinfo{author}{Aubert, J.}, \bibinfo{author}{Fournier, A.}, \bibinfo{year}{2020}.
\newblock \bibinfo{title}{Dynamo-based limit to the extent of a stable layer atop earth’s core}.
\newblock \bibinfo{journal}{Geophysical Journal International} \bibinfo{volume}{222}, \bibinfo{pages}{1433--1448}.
\bibitem[{Gastine et~al.(2015)Gastine, Wicht and Aurnou}]{gastine2015turbulent}
\bibinfo{author}{Gastine, T.}, \bibinfo{author}{Wicht, J.}, \bibinfo{author}{Aurnou, J.M.}, \bibinfo{year}{2015}.
\newblock \bibinfo{title}{Turbulent rayleigh--b{\'e}nard convection in spherical shells}.
\newblock \bibinfo{journal}{Journal of Fluid Mechanics} \bibinfo{volume}{778}, \bibinfo{pages}{721--764}.
\bibitem[{Gilman(1977)}]{Gilman01011977}
\bibinfo{author}{Gilman, P.A.}, \bibinfo{year}{1977}.
\newblock \bibinfo{title}{Nonlinear dynamics of boussinesq convection in a deep rotating spherical shell-i}.
\newblock \bibinfo{journal}{Geophysical \& Astrophysical Fluid Dynamics} \bibinfo{volume}{8}, \bibinfo{pages}{93--135}.
\bibitem[{Gray et~al.(2026)Gray, Guervilly and Sarson}]{GRAY2026107570}
\bibinfo{author}{Gray, M.}, \bibinfo{author}{Guervilly, C.}, \bibinfo{author}{Sarson, G.R.}, \bibinfo{year}{2026}.
\newblock \bibinfo{title}{Influence of rotation on fingering convection in a spherical stably stratified layer}.
\newblock \bibinfo{journal}{Physics of the Earth and Planetary Interiors} \bibinfo{volume}{377}, \bibinfo{pages}{107570}.
\bibitem[{Greenwood et~al.(2021)Greenwood, Davies and Pommier}]{Greenwood2021}
\bibinfo{author}{Greenwood, S.}, \bibinfo{author}{Davies, C.J.}, \bibinfo{author}{Pommier, A.}, \bibinfo{year}{2021}.
\newblock \bibinfo{title}{Influence of thermal stratification on the structure and evolution of the martian core}.
\newblock \bibinfo{journal}{Geophysical Research Letters} \bibinfo{volume}{48}, \bibinfo{pages}{e2021GL095198}.
\bibitem[{Guervilly(2022)}]{guervilly2022fingering}
\bibinfo{author}{Guervilly, C.}, \bibinfo{year}{2022}.
\newblock \bibinfo{title}{Fingering convection in the stably stratified layers of planetary cores}.
\newblock \bibinfo{journal}{Journal of Geophysical Research: Planets} \bibinfo{volume}{127}, \bibinfo{pages}{e2022JE007350}.
\bibitem[{Hirose et~al.(2013)Hirose, Labrosse and Hernlund}]{hirose2013composition}
\bibinfo{author}{Hirose, K.}, \bibinfo{author}{Labrosse, S.}, \bibinfo{author}{Hernlund, J.}, \bibinfo{year}{2013}.
\newblock \bibinfo{title}{Composition and state of the core}.
\newblock \bibinfo{journal}{Annual Review of Earth and Planetary Sciences} \bibinfo{volume}{41}, \bibinfo{pages}{657--691}.
\bibitem[{Hsieh et~al.(2024)Hsieh, Deschamps, Tsao, Yoshino and Lin}]{Hsiehsciadv}
\bibinfo{author}{Hsieh, W.P.}, \bibinfo{author}{Deschamps, F.}, \bibinfo{author}{Tsao, Y.C.}, \bibinfo{author}{Yoshino, T.}, \bibinfo{author}{Lin, J.F.}, \bibinfo{year}{2024}.
\newblock \bibinfo{title}{A thermally conductive martian core and implications for its dynamo cessation}.
\newblock \bibinfo{journal}{Science Advances} \bibinfo{volume}{10}, \bibinfo{pages}{eadk1087}.
\bibitem[{Johnson et~al.(2018)Johnson, Anderson, Korth, Phillips and Philpott}]{Johnson2018}
\bibinfo{author}{Johnson, C.L.}, \bibinfo{author}{Anderson, B.J.}, \bibinfo{author}{Korth, H.}, \bibinfo{author}{Phillips, R.J.}, \bibinfo{author}{Philpott, L.C.}, \bibinfo{year}{2018}.
\newblock \bibinfo{title}{Mercury’s Internal Magnetic Field}. \bibinfo{publisher}{Cambridge University Press}.
\newblock pp. \bibinfo{pages}{114--143}.
\bibitem[{Knibbe and {van Westrenen}(2018)}]{KNIBBE2018147}
\bibinfo{author}{Knibbe, J.S.}, \bibinfo{author}{{van Westrenen}, W.}, \bibinfo{year}{2018}.
\newblock \bibinfo{title}{The thermal evolution of mercury's fe–si core}.
\newblock \bibinfo{journal}{Earth and Planetary Science Letters} \bibinfo{volume}{482}, \bibinfo{pages}{147--159}.
\bibitem[{Kolhey et~al.(2025)Kolhey, Heyner, Wicht, Gastine, Glassmeier and Plaschke}]{Kolhey2025}
\bibinfo{author}{Kolhey, P.}, \bibinfo{author}{Heyner, D.}, \bibinfo{author}{Wicht, J.}, \bibinfo{author}{Gastine, T.}, \bibinfo{author}{Glassmeier, K.H.}, \bibinfo{author}{Plaschke, F.}, \bibinfo{year}{2025}.
\newblock \bibinfo{title}{Dynamo models with a mercury-like magnetic offset dipole}.
\newblock \bibinfo{journal}{Journal of Geophysical Research: Planets} \bibinfo{volume}{130}, \bibinfo{pages}{e2024JE008660}.
\bibitem[{Landeau et~al.(2022)Landeau, Fournier, Nataf, C{\'e}bron and Schaeffer}]{Landeau2022sustaining}
\bibinfo{author}{Landeau, M.}, \bibinfo{author}{Fournier, A.}, \bibinfo{author}{Nataf, H.C.}, \bibinfo{author}{C{\'e}bron, D.}, \bibinfo{author}{Schaeffer, N.}, \bibinfo{year}{2022}.
\newblock \bibinfo{title}{Sustaining earth’s magnetic dynamo}.
\newblock \bibinfo{journal}{Nature Reviews Earth \& Environment} \bibinfo{volume}{3}, \bibinfo{pages}{255--269}.
\bibitem[{Li and Yang(2024)}]{Li_Yang_2024}
\bibinfo{author}{Li, J.}, \bibinfo{author}{Yang, Y.}, \bibinfo{year}{2024}.
\newblock \bibinfo{title}{Double diffusive convection in the diffusive regime with a uniform background shear}.
\newblock \bibinfo{journal}{Journal of Fluid Mechanics} \bibinfo{volume}{993}, \bibinfo{pages}{A14}.
\bibitem[{Loper and Roberts(1981)}]{loper1981study}
\bibinfo{author}{Loper, D.E.}, \bibinfo{author}{Roberts, P.H.}, \bibinfo{year}{1981}.
\newblock \bibinfo{title}{A study of conditions at the inner core boundary of the earth}.
\newblock \bibinfo{journal}{Physics of the Earth and Planetary Interiors} \bibinfo{volume}{24}, \bibinfo{pages}{302--307}.
\bibitem[{Manglik et~al.(2010)Manglik, Wicht and Christensen}]{manglik2010dynamo}
\bibinfo{author}{Manglik, A.}, \bibinfo{author}{Wicht, J.}, \bibinfo{author}{Christensen, U.R.}, \bibinfo{year}{2010}.
\newblock \bibinfo{title}{A dynamo model with double diffusive convection for mercury's core}.
\newblock \bibinfo{journal}{Earth and Planetary Science Letters} \bibinfo{volume}{289}, \bibinfo{pages}{619--628}.
\bibitem[{Mather and Simitev(2021)}]{mather2021regimes}
\bibinfo{author}{Mather, J.F.}, \bibinfo{author}{Simitev, R.D.}, \bibinfo{year}{2021}.
\newblock \bibinfo{title}{Regimes of thermo-compositional convection and related dynamos in rotating spherical shells}.
\newblock \bibinfo{journal}{Geophysical \& Astrophysical Fluid Dynamics} \bibinfo{volume}{115}, \bibinfo{pages}{61--84}.
\bibitem[{Mittelholz and Johnson(2022)}]{Mittelholz}
\bibinfo{author}{Mittelholz, A.}, \bibinfo{author}{Johnson, C.L.}, \bibinfo{year}{2022}.
\newblock \bibinfo{title}{The martian crustal magnetic field}.
\newblock \bibinfo{journal}{Frontiers in Astronomy and Space Sciences} \bibinfo{volume}{Volume 9 - 2022}.
\bibitem[{Monville et~al.(2019)Monville, Vidal, C{\'e}bron and Schaeffer}]{monville2019rotating}
\bibinfo{author}{Monville, R.}, \bibinfo{author}{Vidal, J.}, \bibinfo{author}{C{\'e}bron, D.}, \bibinfo{author}{Schaeffer, N.}, \bibinfo{year}{2019}.
\newblock \bibinfo{title}{Rotating double-diffusive convection in stably stratified planetary cores}.
\newblock \bibinfo{journal}{Geophysical Journal International} \bibinfo{volume}{219}, \bibinfo{pages}{S195--S218}.
\bibitem[{Ouillon et~al.(2020)Ouillon, Edel, Garaud and Meiburg}]{Ouillon_Edel_Garaud_Meiburg_2020}
\bibinfo{author}{Ouillon, R.}, \bibinfo{author}{Edel, P.}, \bibinfo{author}{Garaud, P.}, \bibinfo{author}{Meiburg, E.}, \bibinfo{year}{2020}.
\newblock \bibinfo{title}{Settling-driven large-scale instabilities in double-diffusive convection}.
\newblock \bibinfo{journal}{Journal of Fluid Mechanics} \bibinfo{volume}{901}, \bibinfo{pages}{A12}.
\bibitem[{Radko(2013)}]{radko2013double}
\bibinfo{author}{Radko, T.}, \bibinfo{year}{2013}.
\newblock \bibinfo{title}{Double-diffusive convection}.
\newblock \bibinfo{publisher}{Cambridge University Press}.
\bibitem[{Schaeffer(2013)}]{schaeffer2013efficient}
\bibinfo{author}{Schaeffer, N.}, \bibinfo{year}{2013}.
\newblock \bibinfo{title}{Efficient spherical harmonic transforms aimed at pseudospectral numerical simulations}.
\newblock \bibinfo{journal}{Geochemistry, Geophysics, Geosystems} \bibinfo{volume}{14}, \bibinfo{pages}{751--758}.
\bibitem[{Schaeffer et~al.(2017)Schaeffer, Jault, Nataf and Fournier}]{schaeffer2017turbulent}
\bibinfo{author}{Schaeffer, N.}, \bibinfo{author}{Jault, D.}, \bibinfo{author}{Nataf, H.C.}, \bibinfo{author}{Fournier, A.}, \bibinfo{year}{2017}.
\newblock \bibinfo{title}{{Turbulent geodynamo simulations: a leap towards Earth’s core}}.
\newblock \bibinfo{journal}{Geophysical Journal International} \bibinfo{volume}{211}, \bibinfo{pages}{1--29}.
\bibitem[{Sengupta and Garaud(2018)}]{Sengupta_2018}
\bibinfo{author}{Sengupta, S.}, \bibinfo{author}{Garaud, P.}, \bibinfo{year}{2018}.
\newblock \bibinfo{title}{The effect of rotation on fingering convection in stellar interiors}.
\newblock \bibinfo{journal}{The Astrophysical Journal} \bibinfo{volume}{862}, \bibinfo{pages}{136}.
\bibitem[{Simeonov and Stern(2008)}]{simeonov2008double}
\bibinfo{author}{Simeonov, J.}, \bibinfo{author}{Stern, M.E.}, \bibinfo{year}{2008}.
\newblock \bibinfo{title}{Double-diffusive intrusions in a stable salinity gradient “heated from below”}.
\newblock \bibinfo{journal}{Journal of physical oceanography} \bibinfo{volume}{38}, \bibinfo{pages}{2271--2282}.
\bibitem[{St{\"a}hler et~al.(2021)St{\"a}hler, Khan, Banerdt, Lognonn{\'e}, Giardini, Ceylan, Drilleau, Duran, Garcia, Huang et~al.}]{stahler2021seismic}
\bibinfo{author}{St{\"a}hler, S.C.}, \bibinfo{author}{Khan, A.}, \bibinfo{author}{Banerdt, W.B.}, \bibinfo{author}{Lognonn{\'e}, P.}, \bibinfo{author}{Giardini, D.}, \bibinfo{author}{Ceylan, S.}, \bibinfo{author}{Drilleau, M.}, \bibinfo{author}{Duran, A.C.}, \bibinfo{author}{Garcia, R.F.}, \bibinfo{author}{Huang, Q.}, et~al., \bibinfo{year}{2021}.
\newblock \bibinfo{title}{Seismic detection of the martian core}.
\newblock \bibinfo{journal}{Science} \bibinfo{volume}{373}, \bibinfo{pages}{443--448}.
\bibitem[{Stanley and Mohammadi(2008)}]{STANLEY2008179}
\bibinfo{author}{Stanley, S.}, \bibinfo{author}{Mohammadi, A.}, \bibinfo{year}{2008}.
\newblock \bibinfo{title}{Effects of an outer thin stably stratified layer on planetary dynamos}.
\newblock \bibinfo{journal}{Physics of the Earth and Planetary Interiors} \bibinfo{volume}{168}, \bibinfo{pages}{179--190}.
\bibitem[{Stern(1960)}]{Stern1960}
\bibinfo{author}{Stern, M.E.}, \bibinfo{year}{1960}.
\newblock \bibinfo{title}{The “salt-fountain” and thermohaline convection}.
\newblock \bibinfo{journal}{Tellus} \bibinfo{volume}{12}, \bibinfo{pages}{172--175}.
\bibitem[{Stevenson(2001)}]{stevenson2001mars}
\bibinfo{author}{Stevenson, D.J.}, \bibinfo{year}{2001}.
\newblock \bibinfo{title}{Mars' core and magnetism}.
\newblock \bibinfo{journal}{Nature} \bibinfo{volume}{412}, \bibinfo{pages}{214--219}.
\bibitem[{Stewart et~al.(2007)Stewart, Schmidt, van Westrenen and Liebske}]{Andrewscience}
\bibinfo{author}{Stewart, A.J.}, \bibinfo{author}{Schmidt, M.W.}, \bibinfo{author}{van Westrenen, W.}, \bibinfo{author}{Liebske, C.}, \bibinfo{year}{2007}.
\newblock \bibinfo{title}{Mars: A new core-crystallization regime}.
\newblock \bibinfo{journal}{Science} \bibinfo{volume}{316}, \bibinfo{pages}{1323--1325}.
\bibitem[{Takahashi(2014)}]{takahashi2014double}
\bibinfo{author}{Takahashi, F.}, \bibinfo{year}{2014}.
\newblock \bibinfo{title}{Double diffusive convection in the earth’s core and the morphology of the geomagnetic field}.
\newblock \bibinfo{journal}{Physics of the Earth and Planetary Interiors} \bibinfo{volume}{226}, \bibinfo{pages}{83--87}.
\bibitem[{Takahashi et~al.(2019)Takahashi, Shimizu and Tsunakawa}]{takahashi2019mercury}
\bibinfo{author}{Takahashi, F.}, \bibinfo{author}{Shimizu, H.}, \bibinfo{author}{Tsunakawa, H.}, \bibinfo{year}{2019}.
\newblock \bibinfo{title}{Mercury’s anomalous magnetic field caused by a symmetry-breaking self-regulating dynamo}.
\newblock \bibinfo{journal}{Nature communications} \bibinfo{volume}{10}, \bibinfo{pages}{208}.
\bibitem[{Tassin et~al.(2021)Tassin, Gastine and Fournier}]{tassin2021geomagnetic}
\bibinfo{author}{Tassin, T.}, \bibinfo{author}{Gastine, T.}, \bibinfo{author}{Fournier, A.}, \bibinfo{year}{2021}.
\newblock \bibinfo{title}{Geomagnetic semblance and dipolar--multipolar transition in top-heavy double-diffusive geodynamo models}.
\newblock \bibinfo{journal}{Geophysical Journal International} \bibinfo{volume}{226}, \bibinfo{pages}{1897--1919}.
\bibitem[{Tikoo and Evans(2022)}]{Tikoo2022}
\bibinfo{author}{Tikoo, S.M.}, \bibinfo{author}{Evans, A.J.}, \bibinfo{year}{2022}.
\newblock \bibinfo{title}{Dynamos in the inner solar system}.
\newblock \bibinfo{journal}{Annual Review of Earth and Planetary Sciences} \bibinfo{volume}{50}, \bibinfo{pages}{99--122}.
\bibitem[{Tr{\"u}mper et~al.(2012)Tr{\"u}mper, Breuer and Hansen}]{trumper2012numerical}
\bibinfo{author}{Tr{\"u}mper, T.}, \bibinfo{author}{Breuer, M.}, \bibinfo{author}{Hansen, U.}, \bibinfo{year}{2012}.
\newblock \bibinfo{title}{Numerical study on double-diffusive convection in the earth’s core}.
\newblock \bibinfo{journal}{Physics of the Earth and Planetary Interiors} \bibinfo{volume}{194}, \bibinfo{pages}{55--63}.
\bibitem[{Turner(1985)}]{Turner1985}
\bibinfo{author}{Turner, J.S.}, \bibinfo{year}{1985}.
\newblock \bibinfo{title}{Multicomponent convection}.
\newblock \bibinfo{journal}{Annual Review of Fluid Mechanics} \bibinfo{volume}{17}, \bibinfo{pages}{11--44}.
\bibitem[{Wicht and Sanchez(2019)}]{wicht2019advances}
\bibinfo{author}{Wicht, J.}, \bibinfo{author}{Sanchez, S.}, \bibinfo{year}{2019}.
\newblock \bibinfo{title}{Advances in geodynamo modelling}.
\newblock \bibinfo{journal}{Geophysical \& Astrophysical Fluid Dynamics} \bibinfo{volume}{113}, \bibinfo{pages}{2--50}.
\bibitem[{Wulff et~al.(2025)Wulff, Cao and Aurnou}]{Wulff2025}
\bibinfo{author}{Wulff, P.N.}, \bibinfo{author}{Cao, H.}, \bibinfo{author}{Aurnou, J.M.}, \bibinfo{year}{2025}.
\newblock \bibinfo{title}{On the meaning of the dynamo radius in giant planets with stable layers}.
\newblock \bibinfo{journal}{The Astrophysical Journal} \bibinfo{volume}{992}, \bibinfo{pages}{50}.

\end{thebibliography}

\end{document}